\documentstyle[aasms4]{article}
\def\ubvri{\hbox{$U\!BV\!(RI)_C$}}
\def\ri{\hbox{$R\!-\!I$}}
\font\smcap=cmcsc10

\begin{document}

\title{DISCOVERY OF A LUMINOUS QUASAR IN THE NEARBY UNIVERSE\footnote{Based
on data collected at CNPq/Laborat\'orio Nacional de Astrof\'{\i}sica, Brazil}}

\author{Carlos A. O. Torres\altaffilmark{2}, Germano
R. Quast\altaffilmark{2}, Roger Coziol\altaffilmark{2}}
\author{Francisco Jablonski\altaffilmark{3}, Ramiro de la Reza\altaffilmark{4}, J. R. D. L\'epine\altaffilmark{5}}
\and
\author{J. Greg\'orio-Hetem\altaffilmark{5}}

\altaffiltext{2}{Laborat\'orio Nacional de Astrof\'{\i}sica - LNA/CNPq,
CP 21, 37500--000 Itajub\'a, MG, Brazil} 
\altaffiltext{3}{Divis\~ao de Astrof\'{\i}sica -- INPE/MCT, C.P. 515, 12201-970 S. Jos\'e dos Campos, SP, Brazil}
\altaffiltext{4}{Observat\'orio Nacional, Rua Gal. Jos\'e Cristino, 77, 20921-400 Rio de Janeiro, RJ, Brazil}
\altaffiltext{5}{Instituto Astr\^onomico e Geof\'{\i}sico -- IAG/USP, Av. Miguel St\'efano, 4200, 04301-904 S\~ao Paulo, SP, Brazil}

\begin{abstract}

In the course of the Pico dos Dias survey (PDS), we identified
the stellar like object PDS456 at coordinates 
$\alpha = 17^{\rm h} 28^{\rm m} 19.796^{\rm s}$,  
$\delta = -14^\circ 15' 55.87''$ (epoch 2000), with  
a relatively nearby (z\,$ = 0.184$) and bright (B\,$ = 14.69$) quasar.
Its position at Galactic coordinates l$^{\rm II} = 10^{\circ}.4$, 
b$^{\rm II} = +11^\circ.2$, near the bulge of the Galaxy, may explain why it 
was not detected before. 
The optical spectrum of PDS456 is typical of a luminous quasar, 
showing a broad (FWHM $ \sim 4000$ km s$^{-1}$~) H$\beta$ line,  
very intense Fe{\smcap ii} lines and a weak [\ion{O}{3}]$\lambda5007$ line.
PDS456 is associated to the infrared source IRAS\,17254-1413 with
a 60 $\mu$m infrared luminosity L$_{60} = 3.8 \times 10^{45}$ erg s$^{-1}$. The
relatively flat slopes in the infrared ($\alpha(25,60) = -0.33$ 
and $\alpha(12,25) = -0.78$) and a flat power index in the optical 
(F$_{\nu} \propto \nu^{-0.72}$) may indicate a low dust content.
A good match between the position of PDS456 and the position of the X-ray source 
RXS J172819.3-141600 implies an X-ray luminosity L$_x = 2.8\times 10^{44}$ erg s$^{-1}$. 
The good correlation between the strength of the emission lines in the optical 
and the X-ray luminosity, as well as the steep optical to X-ray index estimated 
($\alpha_{ox} = -1.64$) suggest that PDS456 is radio quiet. A radio survey previously
performed in this region yields an upper limit for radio power at $\sim 5$ GHz 
of $\sim 2.6 \times 10^{30} {\rm erg}\ {\rm s}^{-1} {\rm Hz}^{-1}$. 
We estimate the Galactic reddening in this line--of--sight to be A$_B \simeq 2.0$, implying 
an absolute magnitude M$_B = -26.7$ (using H$_0 = 75$ km s$^{-1}$ Mpc$^{-1}$ and q$_0 = 0$).
In the optical, PDS456 is therefore 1.3 times more luminous than 3C 273 and 
the most luminous quasar in the nearby (z $\leq 0.3$) Universe. 
 
\end{abstract}

 
\section{Introduction} 

The Pico dos Dias survey (PDS) is a systematic search performed at the 
Pico dos Dias Observatory (OPD; operated by LNA/CNPq) to discover young stellar objects. 
Using the Digitized Sky Survey (DSS), we have selected candidates brighter than magnitude 
14 and declination $\delta < +30^\circ$, associated to IRAS 
sources which were chosen following specific color criteria 
(Gregorio-Hetem et al. 1992; Torres et al. 1995). 
High resolution (0.7 \AA) spectra of all the candidates were then taken using 
the Coud\'e spectrograph at the 1.6 m telescope, centering the spectra near H$\alpha$.
\ubvri\ photometry was also performed using FOTRAP (Jablonski et al. 1994), a fast 
photometer installed at the 60 cm Zeiss telescope of the OPD. 
The survey is now complete and final results will soon be published, 
providing information on about 440 new sources.

In this letter, we report the discovery of a new quasar. 
This object, which looks like a 14th magnitude star on the DSS,
is located at coordinates $\alpha = 17^{\rm h} 28^{\rm m} 19.796^{\rm s}$ 
and $\delta = -14^\circ 15' 55.87''$ (astrometric positions, epoch 2000) and
was given the number PDS456 in our catalog. 
Its peculiar position, at Galactic coordinates l$^{\rm II} = 10^\circ.4$, 
b$^{\rm II} = +11^\circ.2$, near the bulge of our Galaxy, may explain 
why this relatively bright quasar has escaped detection until now. 
 
\section {Discovery and observations}

A Coud\'e spectrum of PDS456 was taken on May 12, 1996. On this spectrum 
we distinguished at least three broad emission lines 
suggesting that this object could be an AGN (either a Seyfert or a quasar). 
The narrow wavelength coverage of this high dispersion spectrum
did not allow us, however, to immediately identify the nature of these lines.    
On July 6, 1996 three spectra were obtained with the Cassegrain spectrograph of the OPD.
For this observation, we used a 900 l/mm grating blazed at 5000 \AA, giving 
a resolution of about 3 \AA. One spectrum was taken with five minutes exposure time
and the two others with twenty minutes each. 
The extraction of the spectra was done with the tasks in the APEXTRAC and ONED packages
in IRAF\footnote{IRAF is distributed by National Optical 
Astronomy Observatories, which is operated by the Association of Universities 
for Research in Astronomy, Inc., under contract with the National Science Foundation.}. 
The combination of the three spectra is shown in Figure 1.  

\subsection {Characteristics of the optical spectra}

The Cassegrain spectrum of PDS456 is typical of a luminous quasar. 
We identify the most prominent line with H$\beta$ corresponding
to a redshift of $0.184\pm 0.001$. 
Comparison with the spectrum of PG1700+518 (Wampler 1985) shows very similar characteristics. 
The Fe{\smcap ii} lines are particularly strong and well resolved.
If [O{\smcap iii}]$\lambda$5007 is present, it is much weaker than the 
Fe{\smcap ii} multiplets. 
This is consistent with the anticorrelation found by Boroson \& Green (1992)
between measures of Fe{\smcap ii} and the [\ion{O}{3}].

The analysis of such a complex spectrum is obviously
not the goal of this letter. We did however a preliminary analysis using
in IRAF the routine DEBLEND in SPLOT to verify some hypotheses. 
We decomposed the H$\beta$ line by assuming
two gaussian components: a broad and a narrow one. We did the same analysis
for the broad emission feature at $\sim 5000$ \AA\ assuming that
it was composed of the \ion{Fe}{2} line at $\lambda5018$ \AA\ and 
of the [\ion{O}{3}]$\lambda5007$ line. The results are presented in Table 1. The errors  
were determined by repeating the measures 3 times, varying the level of the continuum  
and the starting position of the lines.  
The results of this analysis are consistent with the hypothesis that  
the FeII lines have the same profile and same widths as H$\beta$ and suggest
also that we do see a weak [\ion{O}{3}]$\lambda5007$. 
The ratio $\log$([\ion{O}{3}]$/{\rm H}\beta) = -0.6$ and the H$\beta$ luminosity 
L(H$\beta$)$= 6.7\times 10^{42} {\rm erg}\ {\rm s}^{-1}$
are in good agreement with the values predicted by the anticorrelation 
found by Steiner (1981).   

\subsection{Photometry and absolute Magnitude of PDS456}

The \ubvri\ magnitudes of PDS456 were obtained on 6 nights in April and May 1997. 
During two nights we observed the quasar twice. The results are summarized in Table 2.
In $V, R$ and $I$ the quasar seems constant, within the measurement errors. In $U$ and
$B$ the variance is somewhat greater, but observations separated 
by three weeks show that the fluxes in these band
do not vary by more than $\sim 10$\%. 

To estimate the reddening in this region of the Galaxy, we used 
the main diffuse interstellar bands present in our spectra (Herbig, 1975, 1993), 
which yields A$_V \sim 1.5$ mag.
This is consistent with the value obtained using the equivalent widths of
the interstellar line Na D 1, assuming that it has multiple components
(Munari \& Zwitter, 1997). 
Our value for the reddening is also consistent with the value A$_B = 1.9$ mag
based on the extinction maps of Burstein and Heiles (1982).  
The extinction--corrected magnitudes and colors are given in column 3 of Table 2. 
We use A$_V = 1.5$ and Seaton's (1979) expression for the reddening in our Galaxy.
The extinction--corrected colors are in relatively good agreement 
with the evolutionary path of colors of quasars at various redshifts 
as derived by Cristiani \& Vio (1990).

The absolute B magnitude of PDS456 was determined using 
the relation (Schmidt \& Green 1983):
${\rm M}_B =  B + 5 - 5 \log({\rm cz}(1+{\rm z}/2)/{\rm H}_0)+{\rm K}-
{\rm A}_{\rm B}$ (using H$_0 = 75$ km s$^{-1}$ Mpc$^{-1}$ and q$_0 = 0$). 
In this expression, the K correction is given by $2.5(1-\alpha)\log(1+{\rm z})$ 
and A$_B$ is the correction for Galactic reddening. 
We used $\alpha = -0.3$ in order to compare with the
absolute magnitudes of the quasars in the V\'eron--Cetty \& V\'eron compendium (1996).
The absolute magnitude of PDS456 is $-26.7$. In figure~2, we compare 
PDS456 with all the quasars in the list of V\'eron--Cetty \& V\'eron.
The absolute magnitudes were reduced to the same cosmology
and corrected for Galactic reddening out of the disk using the expression (Lang 1980):
${\rm A}_B = 0.18[cosec({\rm b}^{\rm II}) - 1] + 0.25$.
A mean value of A$_B = 1.6$ was assumed in the disk ($|{\rm b}^{\rm II}| \leq 5^\circ$). 
In the optical, PDS456 is therefore $\sim 1.3 \times$ more luminous than 3C 273 
(at z $= 0.158$) and the most luminous quasar up to a redshift of $\sim 0.3$.

\subsection{Infrared, X-ray, radio characteristics and spectral energy distribution}

PDS456 is located in the center of the positional error ellipse of 
IRAS\,17254-1413. According to the IRAS Faint Source Catalog  
it was not detected at 100 $\mu$m. The relatively flat
spectral slopes $\alpha(25,60) = -0.33$ and $\alpha(12,25) = -0.78$
may suggest a low dust content. Expressing the 60 $\mu$m luminosity as
(Green et al. 1992) L$_{60} = 4\pi {\rm d}^2 (1+{\rm z})^{-0.1} \nu_{60} f_{60}$,
where d is the distance and $f_{60}$ is the flux in erg cm$^{-2}$ s$^{-1}$
yields L$_{60} = 3.8\times 10^{45}$ erg s$^{-1}$. This is slightly higher than the 
mean value found by Green et al. (1992) for their sample of quasars.  

We also find a good match between the position of PDS456 and the position
of the X-ray source RXS J172819.3-141600 (Voges et al. 1996), which has a count rate of  
0.3 count s$^{-1}$ in the [0.1 - 2.5 kev] energy band. 
Using the conversion factor 1 count s$^{-1} = 1.2\times 10^{-11} {\rm erg}\ {\rm s}^{-1} 
{\rm cm}^{-2}$ as suggested by Alcala (1994) for 1 kev ($\nu = 2.4\times 10^{17}$ Hz), 
we estimate a median flux of 0.0015 mJy. This corresponds to an X-ray 
luminosity L$_X = 2.8\times 10^{44}$ erg s$^{-1}$, which is 
a typical value for quasars. This result is also consistent with the anticorrelation 
between L$_X$ and [O{\smcap iii}]/H$\beta$ 
as found by Grindlay et al. (1980) for radio quiet quasars. 
The ratio $\log({\rm L}_X/{\rm L}_{60}) = -1.1$ is much lower
than the mean for radio quiet quasars and much lower, in particular, than the ratio 
found for 3C 273 ($\log({\rm L}_X/{\rm L}_{60}) = -0.06$; Green et al. 1992).   

The spectral energy distributions of quasars display a wide variety 
of shapes (Barvainis 1990).
In a limited frequency range, it can be described as a power--law in frequencies 
F$_{\nu} \propto \nu^{\alpha}$. 
In Figure 3, we combine the extinction--corrected magnitudes of PDS456 with the IRAS fluxes 
and ROSAT X-ray flux to estimate the spectral 
indexes $\alpha$ in the optical and $\alpha_{ox}$ from the optical to the soft X-rays. 
In the optical, we find $\alpha = -0.72$. This value is significantly
flatter than usually found for quasars and Seyfert 1 galaxies (Edelson \& Malkan, 1986)
which confirms the suggestion given by the slopes in the infrared that 
this quasar is relatively dust free. From the optical to the soft X-rays, we find $\alpha_{ox} = -1.64$. 
This value is consistent with the mean value ($\alpha_{ox} = -1.48$; Laor et al. 1997)
observed for radio quiet quasars.

The characteristics in the optical, the infrared and X-ray as reported
above all suggest that this is a radio quiet quasar. A recent radio survey in
the region of PDS456 did not detect any source to an upper limit of 42 mJy (Griffith et al. 1994). 
At the distance of PDS456 this upper limit corresponds to a 
power level at 4.85 GHz of $2.6 \times 10^{30} {\rm erg\ } {\rm s}^{-1} {\rm Hz}^{-1}$. 
This is two orders of magnitude lower than the power threshold that 
separates radio quiet from radio loud quasars (Woltjer 1990).
 
\section{Discussion}
 
All the studies on the distribution of quasars in space concluded that there is a strong 
evolution of their number density, their luminosity, or both between an epoch corresponding 
approximately to z$\ = 2$ and the present (for a review see Hartwick and Shades 1990). 
The Pure Luminosity Evolution (PLE) model assumes that all the quasars 
were born at the same time and are becoming less luminous at the same rate. 
Alternatively, the Pure Density Evolution (PDE) model 
assumes that the birth rate of the most luminous quasars may have been highest 
in the early Universe and is steadily decreasing, while the birthrate of the least 
luminous ones has roughly been constant.   
In this context, it is intriguing to find a nearby quasar even more luminous
than 3C 273. Indeed, using the HH1 model of Schmidt and Green (1983), 
based on the Bright Quasar Survey (BQS), we find that 
the density $\phi$ of quasars as luminous as 3C 273 in the nearby Universe (z $\leq 0.2$) is
only $\sim 6\times 10^{-11}$ Mpc$^{-3}$. The expected number of such quasars 
in the nearby Universe is $\sim 0.3$. In other words, 3C 273 is an exception.
Therefore the fact that we just discovered another one suggests  
that our understanding of the local luminosity function, as deduced from the BQS, 
is not satisfactory. 

A recent comparison of the Hamburg/ESO survey (HES) with the BQS 
by K\"{o}hler et al. (1997) implies that the surface density of bright 
quasars in the nearby Universe is much higher than previously estimated. 
This result confirms previous claims by Wampler \& Ponz 1985 and Goldshmidt et al. 1992. 
Using the luminosity function of K\"{o}hler et al. and extrapolating 
up to the absolute magnitude of PDS456 we find a density $\log(\phi) = -8.8$ Mpc$^{-3}$. This means
that in a
volume of space including PDS456 we could find as many as 9 quasars 
as luminous as this one. Our discovery supports    
K\"{o}hler et al. claims, which is that the evolution of the most 
luminous quasars in the Universe probably proceeded at a lower rate than usually 
predicted by the PLE model.

\acknowledgments

We would like to thank Carlos Henrique Veiga from the Observat\'orio Nacional who
determined the astrometric coordinates of PDS456 and 
J. E. Steiner for interesting discussions and suggestions.
We would like to thank also the staff at the OPD for their support during our survey.
F. J. and R. C. acknowledge the CNPq for research fellowships. 

\clearpage

\clearpage

\begin{deluxetable}{lcccc}
\footnotesize
\tablecaption{Spectroscopic analysis of PDS456}
\tablewidth{0pt}
\tablehead{
\colhead{feature}& \colhead{$\lambda_{\rm obs.}$} &\colhead{FWHM} & \colhead{EW} & \colhead{F$_{\lambda}$}\nl
\colhead{}&\colhead{}&\colhead{km s$^{-1}$} &\colhead{\AA}&\colhead{$10^{-14}$\ erg cm$^{-2}$ s$^{-1}$}
}
\startdata
H$\beta_{broad}$               &$4849\pm4$    &$3974\pm764$&$21\pm8$    &$28\pm1$   \nl
H$\beta_{narrow}$               &$4862\pm1$    &$1239\pm56 $&$7\pm1 $    &$10.2\pm1$ \nl
[\ion{O}{3}]$\lambda5006.84$   &$5009.8\pm0.1$&$702\pm27$  &$2.1\pm0.3$ &$2.7\pm0.4$\nl
\ion{Fe}{2}$\lambda5018.434$   &$5023.7\pm0.2$&$736\pm39$  &$2.1\pm0.3$ &$2.7\pm0.3$\nl
\enddata
\end{deluxetable}

\begin{deluxetable}{lcc}
\footnotesize
\tablecaption{$UBV(RI)_C$ Photometry of PDS456}
\tablewidth{0pt}
\tablehead{
\colhead{Magnitude/C.I.}   & \colhead{Observed} & \colhead{Dereddened\tablenotemark{a}}}
\startdata
$U$ & $14.12\pm0.06$ & $11.71$ \nl
$B$ & $14.69\pm0.06$ & $12.66$ \nl
$V$ & $14.03\pm0.03$ & $12.44$ \nl
$R$ & $13.58\pm0.03$ & $12.32$ \nl
$I$ & $12.86\pm0.03$ & $11.90$ \nl
\ub & $-0.57\pm0.04$ & $-0.95$ \nl
\bv & $+0.66\pm0.07$ & $+0.22$ \nl
\vr & $+0.45\pm0.05$ & $+0.12$ \nl
\ri & $+0.72\pm0.04$ & $+0.42$ \nl
\enddata
\tablenotetext{a}{Using Seaton's (1979) law, with A$_V=1.5$}
\end{deluxetable}

\clearpage

\figcaption[qso.fig1.eps]{Optical spectra of PDS456. 
The locations of a number of prominent Fe{\smcap ii} lines are indicated.}
\figcaption[qso.fig2.eps]{Comparison of the absolute magnitude of PDS456 with the 
complete list of quasars in the V\'eron-Cetty \& V\'eron compendium (1996). 
The absolute magnitudes were corrected for H$_0 = 75$ and 
for Galactic reddening. PDS456 is $1.3 \times$ more luminous than 3C 273 and the most luminous
quasar up to z$= 0.3$.}
\figcaption[qso.fig3.eps]{Spectral energy distribution of PDS456. The solid lines represent
two power--laws of the form $F_{\nu} \propto \nu^{\alpha}$, with $\alpha = -0.72$ from the far infrared to
the optical and $\alpha_{ox} = -1.64$ from the optical to the soft X-rays.}

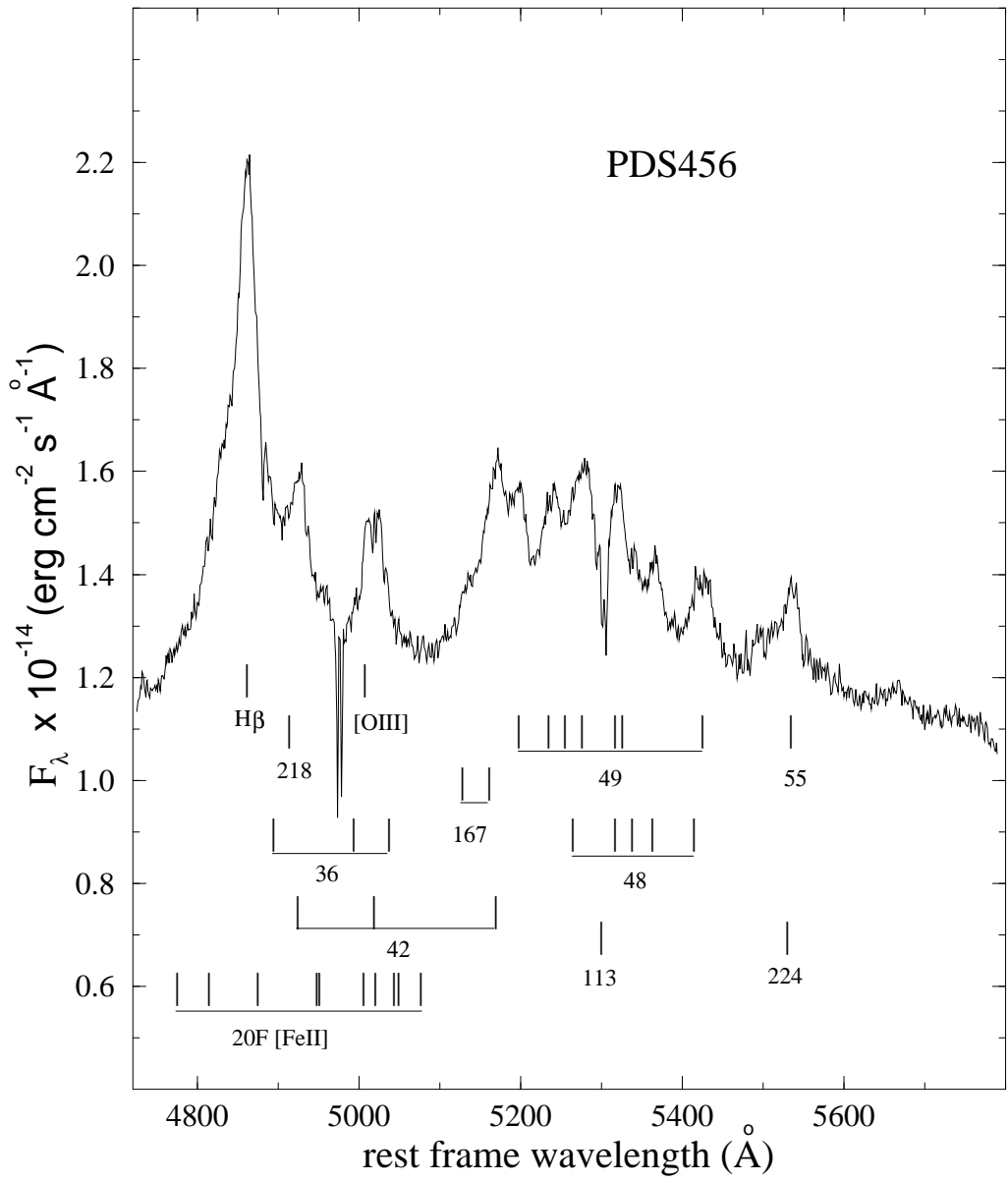
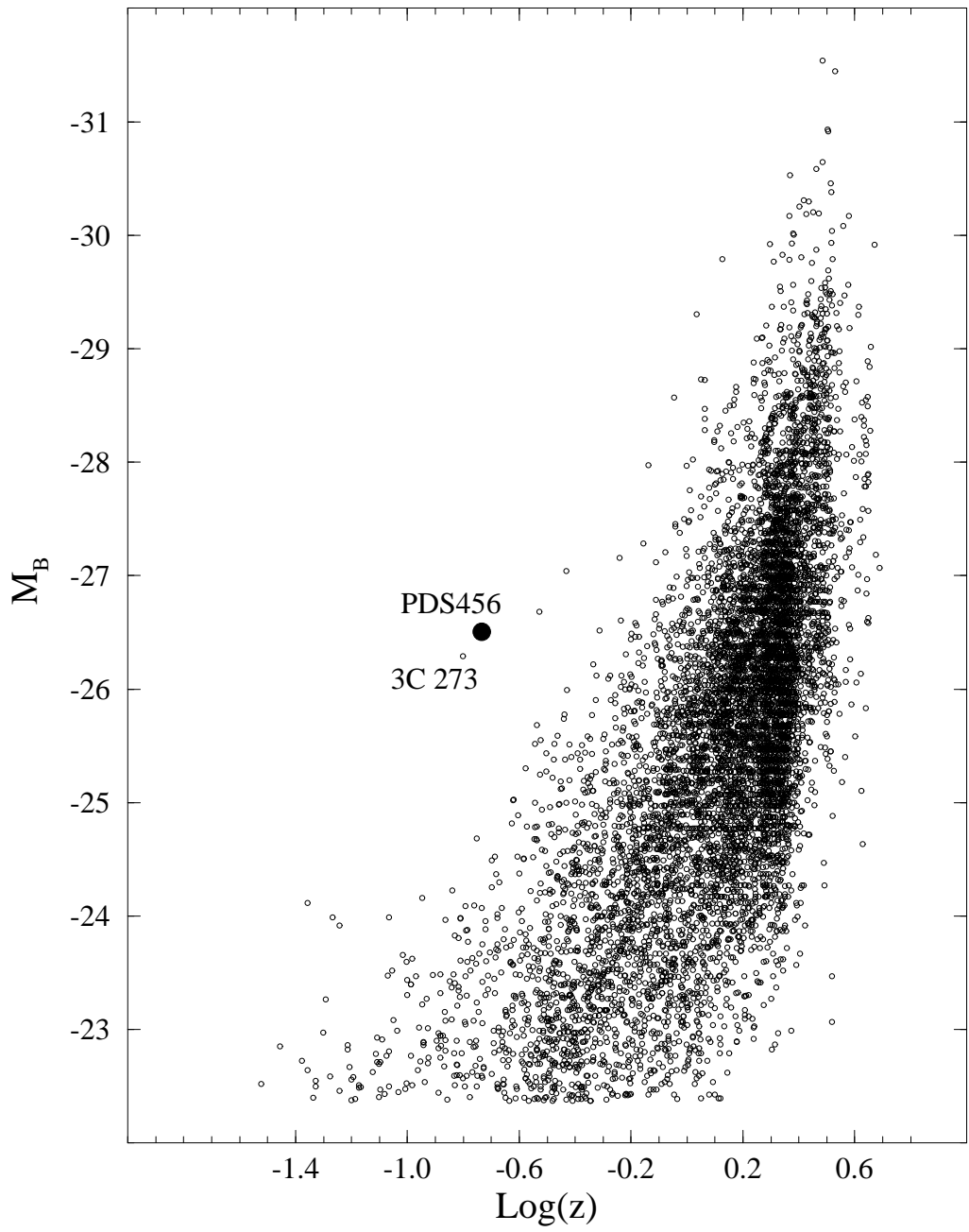
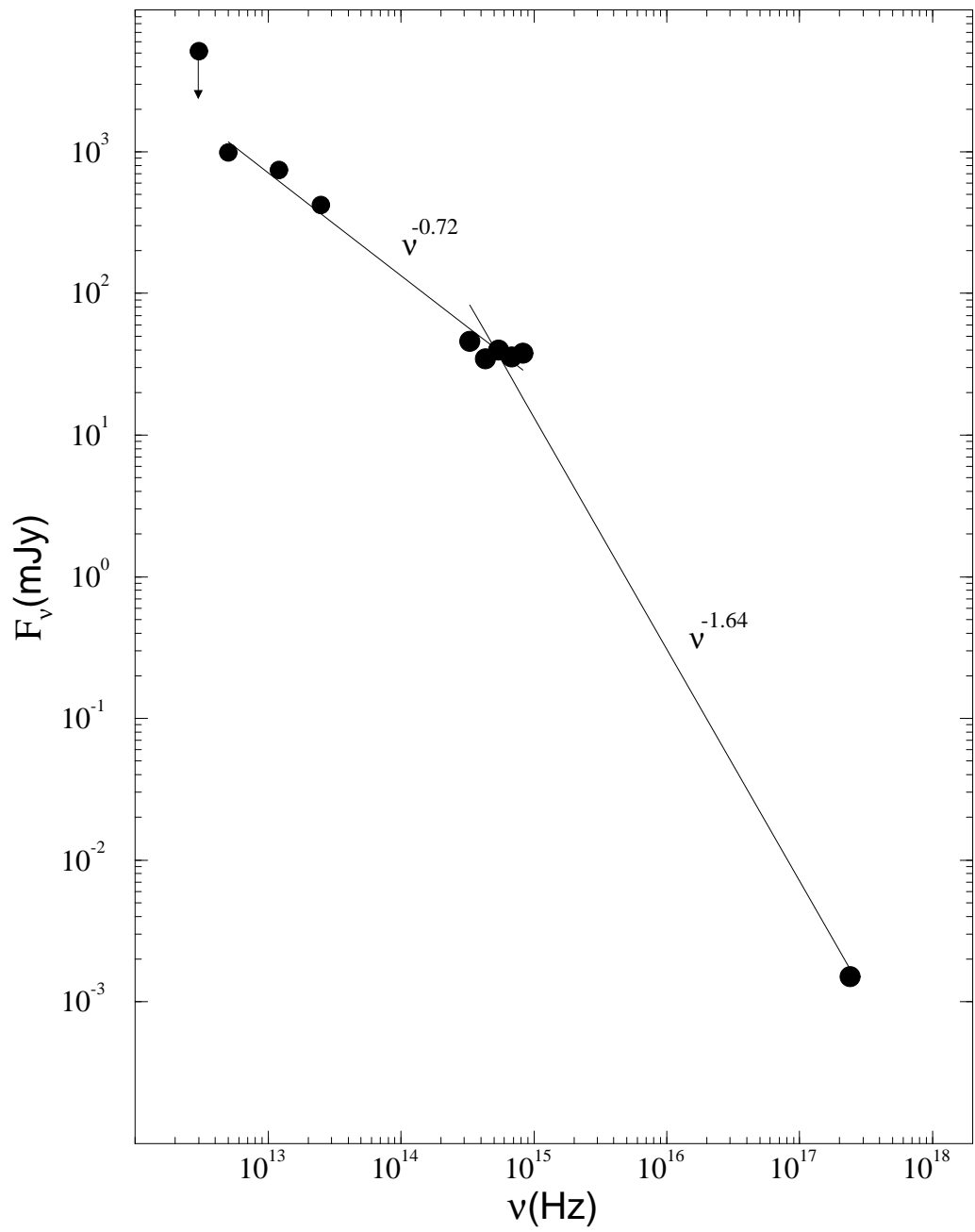
\end{document}